\documentclass[twocolumn,superscriptaddress]{revtex4-2}

\usepackage{graphicx}
\usepackage{color}
\usepackage{soul} 
\usepackage{amsmath}
\usepackage{hyperref}
\usepackage{braket}
\usepackage{mathtools}
\usepackage{lipsum}
\hypersetup{colorlinks=true, citecolor=cyan, urlcolor=blue, linkcolor=blue}
\usepackage[capitalize]{cleveref}
\usepackage{BOONDOX-uprscr}
\usepackage{orcidlink}

\DeclareUnicodeCharacter{2009}{\,}

\newcommand{\RMP}[4]{\textit{#1}, Rev. Mod. Phys. \textbf{#2}, #3 (#4)}
\newcommand{\PR}[4]{\textit{#1}, Phys. Rev. \textbf{#2}, #3 (#4)}
\newcommand{\PRL}[4]{\textit{#1}, Phys. Rev. Lett. \textbf{#2}, #3 (#4)}
\newcommand{\PRR}[4]{\textit{#1}, Phys. Rev. Research. \textbf{#2}, #3 (#4)}
\newcommand{\PRA}[4]{\textit{#1}, Phys. Rev. A \textbf{#2}, #3 (#4)}
\newcommand{\PRB}[4]{\textit{#1}, Phys. Rev. B \textbf{#2}, #3 (#4)}
\newcommand{\PRApplied}[4]{\textit{#1}, Phys. Rev. Appl. \textbf{#2}, #3 (#4)}

\newcommand{\Science}[4]{\textit{#1}, Science \textbf{#2}, #3 (#4)}

\newcommand{\NPhys}[4]{\textit{#1}, Nat. Phys \textbf{#2}, #3 (#4)}
\newcommand{\NComm}[4]{\textit{#1}, Nat. Comm \textbf{#2}, #3 (#4)}

\newcommand{\SciRep}[4]{\textit{#1}, Sci. Rep. \textbf{#2}, #3 (#4)}
\newcommand{\npjQI}[4]{\textit{#1}, npj Quantum Inf. \textbf{#2}, #3 (#4)}

\begin{document}

\title{All-microwave Lamb shift engineering for a fixed frequency multi-level superconducting qubit}

\author{Byoung-moo Ann \orcidlink{0000-0001-6994-0197}}
\email{byoungmoo.ann@gmail.com}
\affiliation{Kavli Institute of Nanoscience, Delft University of Technology, 2628 CJ Delft, The Netherlands} 
\affiliation{Quantum Technology Institute, Korea Research Institute of Standards and Science, 34113 Daejeon, South Korea}
\author{Gary A. Steele}
\affiliation{Kavli Institute of Nanoscience, Delft University of Technology, 2628 CJ Delft, The Netherlands} 
\date{\today}

\begin{abstract}
It is known that the electromagnetic vacuum is responsible for the Lamb shift, which is a crucial phenomenon in quantum electrodynamics (QED). In circuit QED, the readout or bus resonators that are dispersively coupled can result in a significant Lamb shift of the qubit. However, previous approaches or proposals for controlling the Lamb shift in circuit QED demand overheads in circuit designs or non-perturbative renormalization of the system's eigenbases, which can impose formidable limitations.
In this work, we propose and demonstrate an all-microwave method for controlling the Lamb shift of fixed-frequency transmons. We employ the drive-induced longitudinal coupling between the transmon and resonator. By simply using an off-resonant monochromatic drive near the resonator frequency, we can control the net Lamb shift up to ±30 MHz and engineer it to zero with the drive-induced longitudinal coupling without facing the aforementioned challenges. Our work establishes an efficient way of engineering the fundamental effects of the electromagnetic vacuum and provides greater flexibility in non-parametric frequency controls of multilevel systems. 

\end{abstract}

\maketitle
\section*{Introduction}
The rise of modern quantum electrodynamics (QED) was motivated by the need to comprehend the effects of vacuum \cite{Lamb-1, Lamb-2}. One representative phenomenon that accompanied the development of QED is the Lamb shift, which refers to the renormalization of energy levels induced by the electromagnetic fluctuations of the vacuum. Originally, the Lamb shift concerned systems placed in free space. However, the advent of cavity and circuit-QED \cite{cQED0, cQED, Blais-PPA-2006} inspired studies of engineered vacuum. In particular, in circuit-QED, qubits are almost always accompanied by microwave modes in the strong dispersive regime, and Lamb shifts induced by these resonators take significant portions of the bare transition frequency of the qubits \cite{LS-cQED, LS-cQED2, LS-cQED3, LS-cQED4, LS-cQED4-1, LS-cQED4-2, LS-cQED4-3}. 

Thus, controlling the Lamb shift could provide more flexibility in engineering the transition frequencies of superconducting qubits.
In circuit-QED, however, Lamb shift control requires daunting overheads such as flux-tunability \cite{LS-cQED, LS-cQED2, LS-cQED3, LS-cQED4}, voltage biasing \cite{LS-cQED5}, or collective states \cite{LS-cQED6}. Lamb shift can also be controlled without the aforementioned costs using external drivings, as proposed in \cite{Yang-PRA-2010, Jentschura-PRL-2003, Gramich-PRL-2014}. Unfortunately, one cannot avoid mixing among the eigenstates in this manner. Consequently, the properties of the systems will undergo unwanted renormalization \cite{ann_2022}.

In this work, we propose and demonstrate an all-microwave approach for Lamb shift control in a typical circuit-QED configuration comprising a transmon \cite{Koch-PRA-2007} dispersively coupled to a single resonator mode. We introduce strong drive fields off-resonant to both the transmon and resonator, inducing drive-induced longitudinal coupling (DLC). 

This results in state-dependent frequency shifts of the transmon which exist only when the resonator mode is dispersively coupled and therefore can be used to control the Lamb-shift, representing the core-principle of our Lamb shift engineering scheme. We demonstrate large tuning of the Lamb shift $\sim$ 30 MHz while minimizing undesired renormalization of the other properties of the transmon-resonator system.
\begin{figure}
    \centering
    \includegraphics[width=1.05\columnwidth]{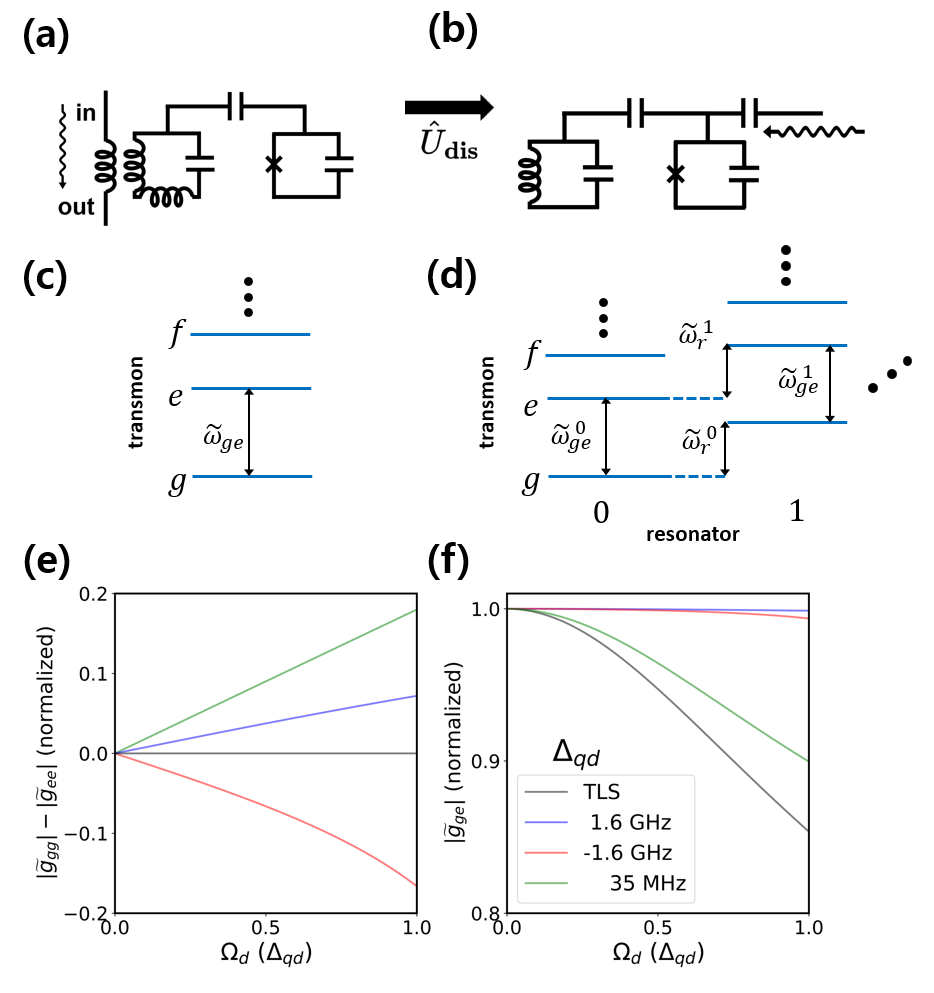}
    \caption{Description of drive-induced Lamb shift engineering. (a) Simplified diagram of circuitry. A transmon is capacitively coupled to a resonator mode. In the experiment, a drive field is inductively applied to the resonator (wavy arrow). (b) Transformed circuitry effectively identical with (a). (c-d) Energy diagram of the effective static Hamiltonians $\hat{K}_q$  and $\hat{K}$. (e-f) Calculated renormalized coupling matrix elements $\widetilde{\mathscr{g}}_{ge},\widetilde{\mathscr{g}}_{gg},\widetilde{\mathscr{g}}_{ee}$ for several transmon–drive field detunings $\Delta_{qd}$ (red, blue, and green). For a two-state (TS) system (black), $\widetilde{\mathscr{g}}_{nm}$ are nearly independent of $\Delta_{qd}$. Drive amplitudes $\Omega_{d}$ in x-axis is normalized by $\Delta_{qd}$. The transmon and resonator's parameters used in the calculation are the same as the experimental values.}
    \label{fig:description}
\end{figure}
\section*{Results}

\subsection*{Theoretical descriptions}
For a dispersively coupled transmon and resonator system, the renormalized interaction in the strong drive limit has been experimentally verified in our previous work \cite{ann_2022}. Unfortuately, the renormalized interaction significantly changes not only the Lamb shift of the transmon, but also other properties such as lifetime, Rabi frequency, and cross-nonlinearity.
In this work, we substantially engineer the Lamb shift while avoiding these unwanted renormalization, which was not dealt with in \cite{ann_2022}.

Fig.~\ref{fig:description}(a) describes an experimental configuration used in this work.
We consider a dispersively coupled transmon and resonator. The drive is inductively applied to the resonator.
In the lab frame, the system Hamiltonian reads 
\begin{equation}
\begin{split}
\label{eq:bare1}
  \hat{H}^{\textbf{(lab)}}  =  ~& \underbrace{4E_C(\hat{N} - N_g)^2 - E_J\cos\hat{\phi}}_{\hat{H}_q} + \underbrace{\omega_r\hat{a}^{\dagger}\hat{a}}_{\hat{
H}_r} \\ & + \underbrace{i\mathscr{g}\hat{N}(\hat{a}-\hat{a}^{\dagger})}_{\hat{H}_I} + \underbrace{\Omega_d^{\textbf{(lab)}} (\hat{a}+\hat{a}^{\dagger}) \sin \omega_d t}_{\hat{H}_d^{\textbf{(lab)}}}.
\end{split}
\end{equation}
$\hat{N}$, $\hat{\phi}$, and $\hat{a}$ refer to cooper-pair number, superconducting phase, and resonator field operator. $E_C$, $E_J$, and $N_g$ are the charging, Josephson energies, and offset cooper-pair numbers of the trasmon. $\Omega_d^{\textbf{(lab)}}$ and $\omega_r$ mean the resonator drive amplitude and frequency. $\mathscr{g}$ is the coupling strength between the transmon and resonator.


To efficiently capture renormalization of the transmon–resonator interaction $\hat{H}_I$, we apply a displacement operator $\hat{U}^{\textbf{(dis)}}=e^{\xi(t)\hat{a}^\dagger-\xi^{*}(t)\hat{a}}$ on $\hat{H}^{\textbf{(lab)}}$.
Here, $\xi(t)=\frac{i\Omega_{d}}{2\Delta_{rd}}e^{-i\omega_{d}t}-\frac{i\Omega_{d}}{2\Sigma_{rd}}e^{i\omega_{d}t}$. $\Delta_{rd}$ and $\Sigma_{rd}$ are $\omega_r - \omega_d$ and $\omega_r + \omega_d$, respectively.
Note that this transformation is only valid when $\Delta_{rd}$ is much larger than the linewidth of the resonator.
Then, the transformed Hamiltonian reads
\begin{equation}
\begin{split}
\label{eq:bare2}
   \hat{H} = ~& \hat{U}^{\textbf{(dis)}}[\hat{H}^{\textbf{(lab)}}-i\partial_t]\hat{U}^{\textbf{(dis)}\dagger} \\ = ~&\hat{H}_q + \hat{H}_I + \hat{H}_r  + \underbrace{\Omega_d\hat{N}\cos\omega_d t}_{\hat{H}_d}.
\end{split}
\end{equation}
While eliminating $\hat{H}_d^{\textbf{(lab)}}$, we obtain a transmon drive $\hat{H}_d$, and $\Omega_d$ therein is $\mathscr{g}(\frac{\Omega_d^{\textbf{(lab)}}}{{\Delta}_{rd}}-\frac{\Omega_d^{\textbf{(lab)}}}{\Sigma_{rd}})$. An equivalent circuit configuration is given in Fig.~\ref{fig:description}(b).

We introduce unitary transformations $\hat{U}_q$ and $\hat{U}$, which transform $\hat{H}_q + \hat{H}_d$ and $\hat{H}$ to effective static Hamiltonian $\hat{K}_q$ and $\hat{K}$ respectively \cite{ann_2022}. We depict the energy levels of $\hat{K}_q$ and $\hat{K}$ in Fig.~\ref{fig:description}(c–d). We define $\widetilde{\omega}_{nm}$ the transition frequency between $n$-th and $m$-th states of $\hat{K}_q$. We also define $\widetilde{\omega}_{nm}^{k}$ ($\widetilde{\omega}_{r}^{l}$), which refers to the transmon (resonator) transition frequency when the resonator (transmon) is in the $k$-th ($l$-th) state. 
To efficiently distinguish the transmon and resonator states, we label the lowest four states of the transmon by $g$, $e$, $f$, and $d$, respectively.

The difference between $\hat{K}_q$ and $\hat{K}$ is originated from the interaction between the transmon and resonator. Particularly, the discrepancy between $\widetilde{\omega}_{nm}^0$ and $\widetilde{\omega}_{nm}$ can be interpreted as a transmon frequency shift when the resonator is in vacuum. Therefore, we can define renormalized Lamb shift $\widetilde{L}_{nm} = \widetilde{\omega}_{nm}^0 - \widetilde{\omega}_{nm}$, and resonator frequency pulling $\widetilde{P}=\widetilde{\omega}_r^g-\omega_r$.
All $\widetilde{\omega}_{nm}$, $\widetilde{\omega}_{nm}^{k}$, $\widetilde{L}_{nm}$, and $\widetilde{P}$ are adiabatically connected to ${\omega}_{nm}$, ${\omega}_{nm}^{k}$, ${L}_{nm}$, and ${P}$  with $\Omega_{d}\rightarrow0$.
We further define AC Stark shift of the transmon $\delta\omega_{nm}=\widetilde{\omega}_{nm}-\omega_{nm}$. We also define $\delta\omega_{nm}^{k}=\widetilde{\omega}_{nm}^{k}-\omega_{nm}^{k}$. For far off-resonant drives, $\delta\omega_{nm} \approx \delta\omega_{nm}^{k}$ is satisfied since the interplay between AC Stark and Lamb shift is negligible.

To gain an intuition of how the transmon–resonator interaction accounts for the difference in $\widetilde{\omega}_{nm}$ and $\widetilde{\omega}_{nm}^{0}$, it is useful to define the renormalized interaction Hamiltonian \cite{ann_2022}
\begin{equation}
\begin{split}
\label{eq:dressed-dipole}
   & \hat{\widetilde{H}}_I =  i\mathscr{g}[\hat{U}_q\hat{N}\hat{U}_q^\dagger](\hat{a} - \hat{a}^\dagger) \\ \cong  &i\sum_{n,m}\widetilde{\mathscr{g}}_{nm}(e^{i(n-m+1)\omega_d t}-e^{i(n-m-1)\omega_d t})\ket{n}\bra{m}(\hat{a} - \hat{a}^\dagger).
\end{split}
\end{equation}
Here, $\ket{n}$ is the eigenstate of $\hat{H}_{q}$.
For the discussion later, we define $\hat{\widetilde{H}}_{\textsc{DLC}}$, the renormalized interaction Hamiltonian containing only drive-induced longitudinal coupling (DLC) 
\begin{equation}
\begin{split}
\label{eq:dressed-dipole-l}
   & \hat{\widetilde{H}}_{\textsc{DLC}} =  i\sum_{n}\widetilde{\mathscr{g}}_{nn}(e^{i\omega_d t}-e^{-i\omega_d t})\ket{n}\bra{n}(\hat{a} - \hat{a}^\dagger).
\end{split}
\end{equation}

For far off-resonant drives, the magnitudes of static components ($n-m=\pm1$) in Eq.~\ref{eq:dressed-dipole} remain nearly invariant. Also, the magnitudes of off-diagonal dynamical components ($n\neq m$ and $n-m\neq\pm1$) are much smaller compared to those of the static components.
In this work, we focus on the DLC terms in Eq.~\ref{eq:dressed-dipole-l}, which in turn significantly contribute to $\widetilde{L}_{nm}$.

In Fig.\ref{fig:description}(e-f), we theoretically calculate some elements of static  ($\widetilde{\mathscr{g}}_{ge}$) and DLC terms ($\widetilde{\mathscr{g}}_{gg, ee}$) for several $\Delta_{qd} = \omega_{ge} - \omega_d$ and $\Omega_d$ Based on \cite{ann_2022}.
These mainly determine $\widetilde{L}_{ge}$.
The parameters used in the calculation are the same as the experimental values.
In Fig.\ref{fig:description}(e), we observe the discrepancy between $|\widetilde{\mathscr{g}}_{gg}|$ and $|\widetilde{\mathscr{g}}_{ee}|$ for both far-off-resonant (red and blue) and near-resonant (green) drive fields.
For two-state (TS) systems, $|\widetilde{\mathscr{g}}_{gg}| = |\widetilde{\mathscr{g}}_{ee}|$ always holds. 
Fig.\ref{fig:description}(f) presents $|\widetilde{\mathscr{g}}_{ge}| (= |\widetilde{\mathscr{g}}_{eg}|)$.
As we can confirm in Fig.\ref{fig:description}(f), near-resonant driving significantly renormalizes $\widetilde{\mathscr{g}}_{ge}$.
For far-off-resonant driving, the static components remain nearly the same. In addition, the magnitude of other off-diagonal dynamical terms are negligible (not present in Fig.\ref{fig:description}(f)). Therefore, the transverse part in the renormalized interaction Hamiltonian can be approximated to $\hat{H}_{I}$.

Eventually, taking only the static and DLC components into consideration, we can approximate $\widetilde{L}_{n,n+1}$ by 
\begin{equation}
\begin{split}
\label{eq:approx}
   & \widetilde{L}_{n,n+1} \approx \frac{|\widetilde{\mathscr{g}}_{n,n+1}|^2}{\widetilde{\omega}_{n,n+1}-\omega_r} + \frac{|\widetilde{\mathscr{g}}_{nn}|^2 - |\widetilde{\mathscr{g}}_{n+1,n+1}|^2}{{\omega}_{d} - {\omega}_{r}}.
\end{split}
\end{equation}
Eq.~\ref{eq:approx} provides a rough estimation of $\widetilde{L}_{n,n+1}$ when $ 
|\widetilde{\mathscr{g}}_{n,n+1}| \ll |\widetilde{\omega}_{ge}-\omega_r|$ and $|\widetilde{\mathscr{g}}_{nn}| , |\widetilde{\mathscr{g}}_{n+1,n+1}|  \ll |{\omega}_{d} - {\omega}_{r}|$ are satisfied.
The first term describes the Lamb shift induced by the static components in Eq.~\ref{eq:dressed-dipole}. The second term corresponds to the Lamb shift induced by DLC.
When $\omega_d$ is closed to $\omega_r$, the DLC-induced Lamb shift can contribute significantly to $\widetilde{L}_{n,n+1}$ keeping $\widetilde{\mathscr{g}}_{n,n+1} \approx {\mathscr{g}}_{n,n+1}$.

This scheme is not possible for two-state system for $|\widetilde{\mathscr{g}}_{gg}| = |\widetilde{\mathscr{g}}_{ee}|$.
In \textbf{Supplementary Note 2}, we generalize the theoretical description in this subsection to arbitrary multi-level systems coupled to resonator modes based on Floquet formalism.

\subsection*{Experimental conditions}
We obtain the experimental data from two cooldowns due to an accidental interruption in the experiment caused by a technical issue.
The circuit parameters for each round are distinguished by unbracketed (1st) and bracketed values (2nd).
The data in Fig.~\ref{fig2} is obtained in the first round. Fig.~\ref{fig3} and Fig.~\ref{fig4} are obtained from the data in the second round.
From the pulsed qubit spectroscopy, we obtain $\omega_{ge}^{0}/2\pi\approx5.901 (5.867)$GHz, $\omega_{ef}^{0}/2\pi\approx5.749 (5.715)$GHz, $\omega_{fd}^{0}/2\pi\approx5.587 (5.553)$GHz, and $\omega_r^g/2\pi\approx 4.290 (4.289)$GHz.
We also obtain $\omega_r/2\pi\approx4.335 (4.335)$GHz by driving the transmon to unconfined states \cite{unconfined}.
Based on these, we extract bare qubit parameters and coupling, $\omega_{ge}/2\pi\approx5.869 (5.835)$GHz, $\omega_{ef}/2\pi\approx5.708 (5.676)$GHz, $\omega_{fd}/2\pi\approx5.539 (5.510)$GHz, and $\mathscr{g}/2\pi\approx248(245)$MHz.
The extracted parameters are consistent with the observed self and cross-nonlinearity, $A=\omega_{ge}^{0}-\omega_{ef}^{0}\approx2\pi\times152(150)$MHz and $\chi=\omega_{r}^{g}-\omega_{r}^{e}\approx2\pi\times5.8(6.0)$MHz, respectively. Please see \textbf{Supplementary Note 1, Supplementary Table 1 and 2} for detailed information on system parameters and variables.

\subsection*{Resolving drive-induced longitudinal coupling}

Experimentally verifying the existence of drive-induced longitudinal coupling (DLC) is non-trivial. Both DLC and AC Stark shifts yields $\delta\omega_{nm}^{0}\sim O(\Omega_{d}^2)$, and thus, one cannot distinguish them just simply measuring the changes in $\omega_{nm}^{0}$ without independent calibration of $\Omega_{d}$. Instead, we investigate the ratios among $\delta\omega_{nm}^{0}$ to identify the DLC. We introduce the following dimensionless quantities.
\begin{equation}
\begin{split}
\label{eq:eta}
    &\eta_{ef}^{n}=\frac{1}{2}\delta\omega_{gf}^{n}/\delta\omega_{ge}^{n}|_{\Omega_{d}\rightarrow0},\\
    &\eta_{ed}^{n}=\frac{1}{3}\delta\omega_{gd}^{n}/\delta\omega_{ge}^{n}|_{\Omega_{d}\rightarrow0}.
\end{split}
\end{equation}
We will compare experimentally obtained $\eta$ to the theory with and without considering DLC, and thereby verify the effects of the DLC. Note that finding experimental $\eta$ does not demand calibrating $\Omega_{d}$ since it is independent of $\Omega_{d}$.

\begin{figure}
    \centering
    \includegraphics[width=1.05\columnwidth]{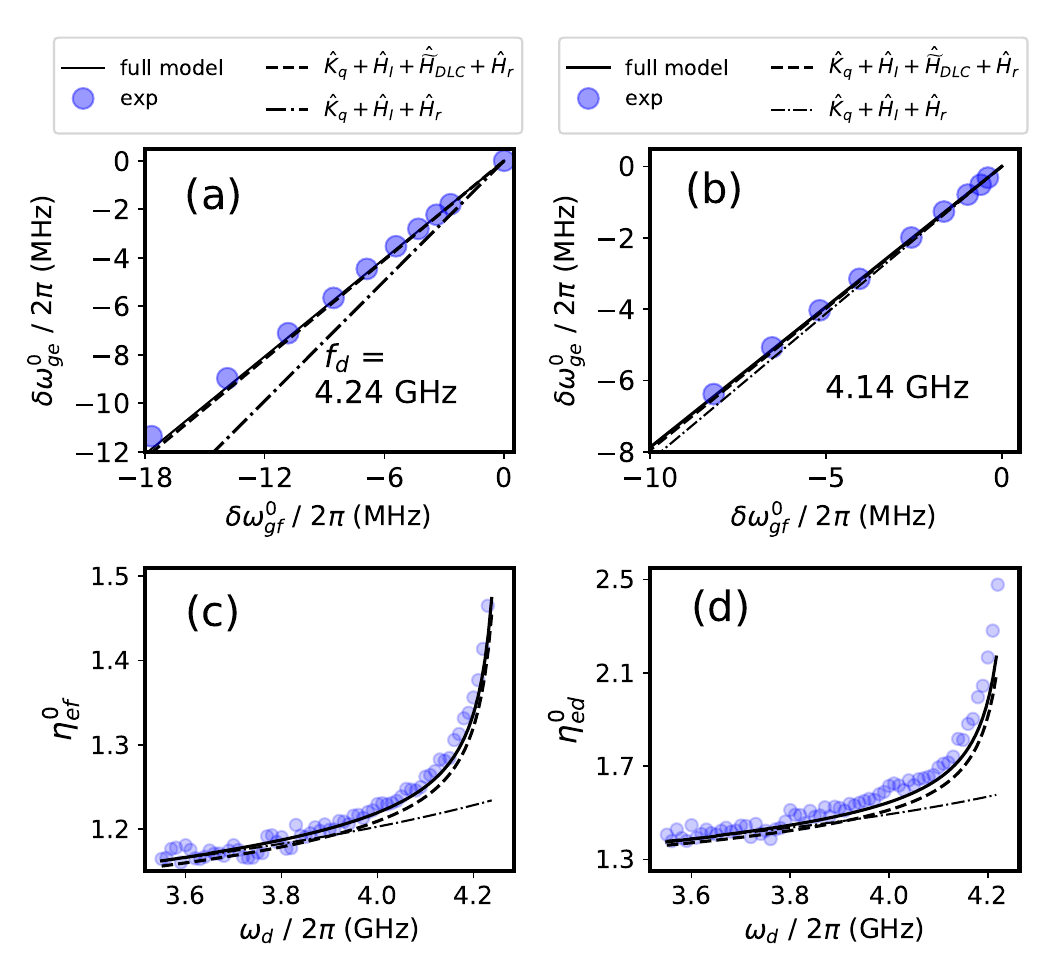}
    \caption{Identifying drive-induced longitudinal coupling (DLC) from multi-level spectroscopy. We investigate drive frequency $\omega_d$ near $\omega_r^g$. 
    Circles denote experimental data. Lines indicate theoretical calculation based on the corresponding Hamiltonian models in legend. (a-b) We plot the frequency shifts in $ge$ transition ($\delta\omega_{ge}^0$) with respect to that of $gf$ transition ($\delta\omega_{gf}^0$) for $\omega_d/2\pi=f_d=4.24$ GHz and $\omega_d/2\pi=f_d=4.14$ GHz, respectively. (c-d) We plot dimensionless quantities $\eta_{ef}^0=\frac{1}{2}\delta\omega_{gf}^{n}/\delta\omega_{ge}^{n}|_{\Omega_{d}\rightarrow0}$ and $\eta_{ed}^0=\frac{1}{2}\delta\omega_{gd}^{n}/\delta\omega_{ge}^{n}|_{\Omega_{d}\rightarrow0}$, while sweeping $\omega_d$. Errors are less than the size of symbols, and thus not presented in the plots. The errors are statistical and originated when extracting $\delta\omega^0$ from data. 
    }
    \label{fig2}
\end{figure}

In Fig.~\ref{fig2}, we measure both $\eta_{ef}^{0}$ and $\eta_{ed}^{0}$ from multi-level spectroscopy \cite{multi}. Please see \textbf{Supplementary Note 3} for details on the experimental methods.
In Fig.~\ref{fig2}(a-b), we present the observed $\delta\omega_{ge}^0$  with respect to $\delta\omega_{gf}^0$ (circles) for two different drive frequencies near $\omega_r^g$, 4.24 GHz (a) and 4.14 GHz (b), respectively.
We choose $\omega_d$ to be close enough to $\omega_r$ since the effects of the DLC scale linearly with $1/(\omega_d - \omega_r)$ as shown in Eq.~\ref{eq:approx}.
We confirm linear correlations among experimentally observed $\delta\omega_{ge,gf,gd}^{0}$ for $\delta\omega_{ge,gf,gd}^{0}/ 2\pi \lesssim 10$MHz as seen in Fig.~\ref{fig2}(a-b).
In Fig.~\ref{fig2}(c-d), we sweep $\omega_d$ from 3.55 GHz to 4.25 GHz and present corresponding $\eta_{ef}^0$ and $\eta_{ed}^0$ from the experiments (circles).

The solid, single-dashed, and dot-dashed lines refer to the theoretical calculations based on $\hat{K}_q + \hat{\widetilde{H}}_I + \hat{H}_r$, $\hat{K}_q +\hat{H}_{I} + \hat{\widetilde{H}}_{\textsc{DLC}} + \hat{H}_r$, and $\hat{K}_q+\hat{H}_{I} + \hat{H}_r$, respectively.
We apply Floquet theory \cite{Floquet_1,Floquet_2} to the above Hamiltonians and calculate the theoretical values. The calculations are numerically done by QuTip \cite{Qutip1,Qutip2}.

The first model presents a full description of the driven system, which excellently explains the experimental data.
The second and third models differs only by a term $\hat{\widetilde{H}}_{\textsc{DLC}}$.
Therefore, the disagreements between these models can be interpreted as the effects from the DLC.
The breakdown of the dot-dashed lines in Fig.~\ref{fig2}(a-d), and the excellent consistency among the experiment, the solid and single-dashed lines indicate clear evidences for the DLC.
As expected from Eq.~\ref{eq:approx}, we can confirm that the DLC effect is larger with smaller $|\omega_d-\omega_r|$ in Fig.~\ref{fig2}(a-b).
Such tendency is also clearly confirmed in Fig.~\ref{fig2}(c-d).

From the investigation of this section, we conclude that calibrating $\Omega_d$ cannot be precisely achieved only using AC Stark shift theory since the DLC should take a significant portion of the frequency shifts. In the following section, we use a more rigorous approach to find $\Omega_d$, and thereby extract the Lamb shifts at arbitrary drives.

\subsection*{Lamb shift renormalization at arbitrary drive strengths}
\begin{figure}
    \centering
    \includegraphics[width=1\columnwidth]{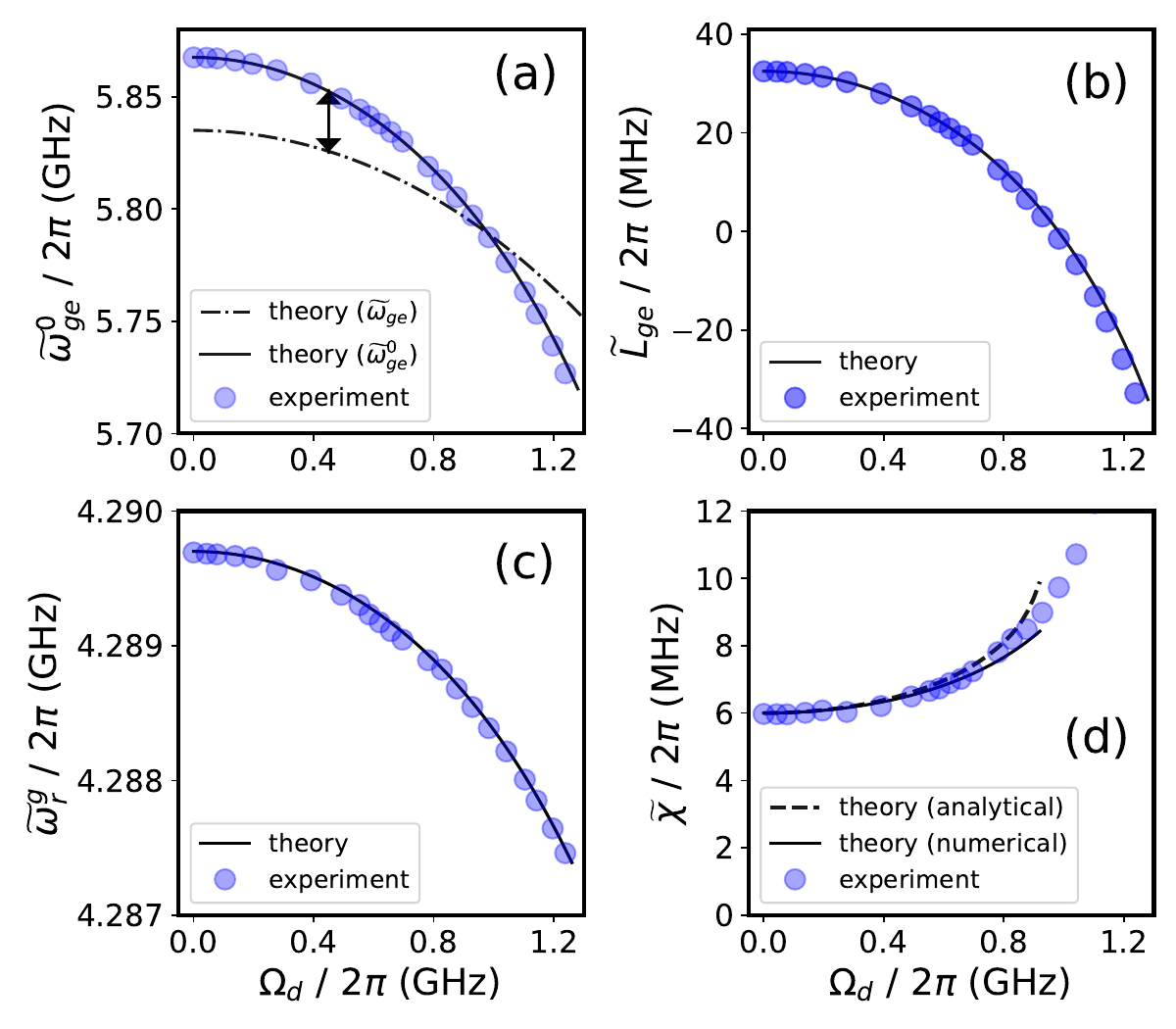}
    \caption{Lamb shift and other renormalized quantities with respect to drive amplitude $\Omega_{d}$. Drive frequency $\omega_d/2\pi$ is $4.2$ GHz for all cases. Circles and lines denote experimental data and theoretical calculation, respectively. We plot the renormalized transmon transition frequency ($\widetilde{\omega}_{ge}^{0}$ and $\widetilde{\omega}_{ge}$) in (a), Lamb shift ($\widetilde{L}_{ge}$) in (b), resonator frequency ($\widetilde{\omega}_r^g$) in (c), and cross-nonlinearity ($\widetilde{\chi}$) in (d). Errors are less than the size of symbols, and thus not presented in the plots. Errors are statistical and originated when extracting $\delta\omega^0$ from data. }
    \label{fig3}
\end{figure}

In the previous section, we have proven the existence of DLC effects, and thereby learned employing AC Stark shift theory alone is an inappropriate approach to calibrate $\Omega_{d}$. From now on, we use Eq.~\ref{eq:bare2}, including the resonator and interaction terms, to obtain $\Omega_{d}$ in the experiment. We then quantify the renormalized Lamb shift at arbitrary $\Omega_{d}$. We cross-check our quantification from the shifts in the resonator frequency and cross-nonlinearity. Note that the AC Stark shift alone cannot explain these shifts simultaneously.

Fig.~\ref{fig3} present experimentally observed $\widetilde{\omega}_{ge}^0$, $\widetilde{\omega}_r^g$, and $\widetilde{\chi}$ (circles) for $\omega_d/2\pi=4.2$ GHz.
We first obtain the conversion factor $\mu(\omega_d)$ that satisfies $\mu(\omega_d) \sqrt{P_d}=\Omega_{d}$, where $P_d$ indicates the driving power measured at the signal generator.
We set $\mu=138.9$, with which all quantities are simultaneously explained by the theories.

In Fig.~\ref{fig3}(a), we compare experimentally observed $\widetilde{\omega}_{ge}^0$ to theoretical expectation (solid line). 
For a comparison, we plot the $\widetilde{\omega}_{ge}$ (dot-dashed line) theoretically calculated based on $\hat{K}_q$.
An arrow indicates $\widetilde{L}_{ge} = \widetilde{\omega}_{ge}^0 - \widetilde{\omega}_{ge}$. There is a crossing between the data and dot-dashed line, which means the sign of $\widetilde{L}_{ge}$ is flipped at that drive amplitude. 
In Fig.~\ref{fig3}(b–c), we plot experimentally observed $\widetilde{L}_{ge}$ and $\widetilde{\omega}_r^g$ with the theoretical expectation (lines). $\widetilde{L}_{ge}$ varies from $32$ to $-30$ MHz.
The changes in the resonator frequency pulling $\widetilde{P} = \widetilde{\omega}_r^g -{\omega}_r$ is relatively less than those of $\widetilde{L}_{ge}$. 
All the theoretical calculations in (a–c) are based on Floquet theory and numerically performed by QuTiP \cite{Qutip1,Qutip2}.

We present the renormalized cross-nonlinearities ($\widetilde{\chi}$) of the driven transmon–resonator system in Fig.~\ref{fig3}(d). The circles and lines indicate the experimental and theoretical calculation, respectively.
We investigate the origin of $\Omega_{d}$ dependence of $\widetilde{\chi}$.
In the analytical theory (dashed line), we use the perturbative calculation $\widetilde{\chi} \approx \widetilde{\mathscr{g}}_{ge}^2\widetilde{A}/(\widetilde{\omega}_{ge}^0 - \widetilde{\omega}_{r}^0 - \widetilde{A})$ \cite{Koch-PRA-2007}, and use the approximation $\widetilde{\mathscr{g}}_{ge}\approx{\mathscr{g}}_{ge}$.
Here, $\widetilde{A}$ is the renormalized self-nonlinearity, $\widetilde{\omega}_{ge}^{0}-\widetilde{\omega}_{ef}^{0}$. We do not make any approximation on $\widetilde{A}$ in the analytical calculation.
The analytical theory is consistent with the experimental data as well as the numerical calculation based on Floquet theory (solid line). Therefore, we can conclude that the approximation $\widetilde{\mathscr{g}}_{ge}\approx{\mathscr{g}}_{ge}$ is satisfied. 
The disagreement between solid and dashed lines at large $\Omega_{d}$ in Fig.~\ref{fig3}(d) can be attributed to undesired sideband transitions between the transmon and resonator. See \textbf{Supplementary Note 3} for more detailed discussion.

\subsection*{Drive-induced dephasing}

\begin{figure}
    \centering
    \includegraphics[width=1\columnwidth]{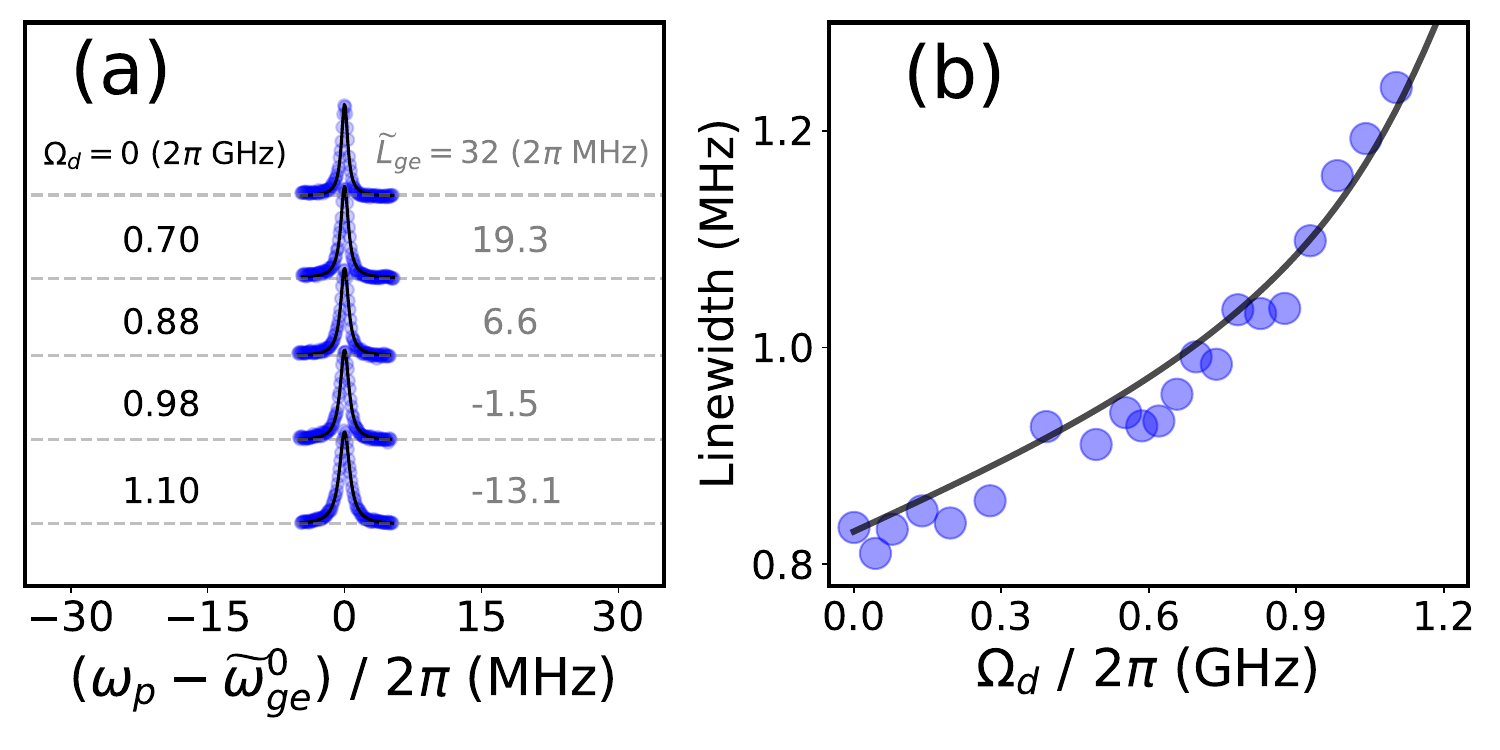}
    \caption{Linewidth broadening by the drive-induced dephasing.  Drive frequency $\omega_d/2\pi$ is set by 4.2 GHz. (a) The transmon's two-tone spectroscopy data with respect to various $\Omega_{d}$. $\omega_p$ refers to probe frequency. Corresponding renormalized Lamb shift $\widetilde{L}_{ge}$ are also presented. Circles and lines denote data and Lorentzian fits. (b) Extracted linewidths with respect to $\Omega_{d}$. Line is obtained by theoretical model. The linewidth broadening is originated by the finite lifetime of the resonator. Errors are less than the size of symbols, and thus not presented in the plots. The errors in (b) are statistical and originated when extracting the linewidth from (a). }
    \label{fig4}
\end{figure}

In Fig.~\ref{fig4}, we investigate how the transmon's linewidth varies while engineering $\widetilde{L}_{ge}$ from 32 to $-$30 MHz.
Fig.~\ref{fig4}(a) shows two-tone spectroscopy of $g\rightarrow e$ transition for various $\Omega_{d}$. Corresponding $\widetilde{L}_{ge}$ is also presented beside.
We obtain $\Gamma_1^{q}\approx 1$ MHz and $\Gamma_{\phi}^{q}\approx 2$ MHz from time-domain measurement, where $\Gamma_1^{q}$ and $\Gamma_{\phi}^{q}$ are energy relaxation and pure dephasing rates of the transmon.
Corresponding linewidth in two-tone spectroscopy is approximately 830 kHz without probe power broadening and measurement-induced dephasing \cite{two-tone,two-tone2}.
We also obtain the similar linewidth from two-tone spectroscopy in the experiment, when the calibrated pump strength is approximately 110 kHz, and measurement photon number is far less than unity.
There are almost no qualitative changes in the spectrum presented in Fig.~\ref{fig4}(a) with increasing $\Omega_{d}$. However, we notice the linewidth increases by a significant amount.
Fig.~\ref{fig4}(b) shows the extracted linewidth from Lorentzian fitting (circles).
We name such effect drive-induced dephasing (DID) in this paper. 

We reveal that the cooperative effects from the driving and finite resonator lifetime can explain the linewidth broadening. The amount of DID is defined by $\Gamma_{\phi,\textbf{DID}}^{q}$.
The same phenomenon is also theoretically predicted in \cite{DID}, but has been rarely demonstrated experimentally.
Based on Eq.33 of \cite{DID}, we obtain the approximated form of $\Gamma_{\phi,\textbf{DID}}^{q}$
\begin{equation}
\begin{split}
    \Gamma_{\phi,\textbf{DID}}^{q} \approx \frac{\sqrt{\widetilde{A}\widetilde{\chi}}}{2\widetilde{\Delta}_{rd}}\frac{\Omega_{d}}{2\widetilde{\Delta}_{qd}} \times \Gamma_{1}^{r}(\omega_d).
\label{dlc-dephasing}
\end{split}
\end{equation}
$\widetilde{\Delta}_{qd}$ and $\widetilde{\Delta}_{rd}$ are given by $\widetilde{\omega}_{ge}^0-\omega_d$ and $\widetilde{\omega}_{r}^g-\omega_d$, respectively. 
$\Gamma_1^{r}(\omega)$ is the resonator–bath coupling.
We have $\Gamma_1^{r}(\omega_r^g)=13.47$ MHz from the resonator decay rate, which is mainly accounted for by the external coupling to the feedline.
The theory curve in Fig.~\ref{fig4}(b) is based on Eq.~\ref{dlc-dephasing}.
$\Gamma_1^{r}(\omega_d)$ is determined by some unknown factors such as the cable resonances of feedlines, and empirically known slowly varying over a few hundreds MHz frequency scale. 
Thus, we set $\Gamma_1^{r}(\omega_d)$ as a free-fitting parameter and obtain the value of $(0.83 \pm 0.05) \times \Gamma_1^{r}(\omega_r^g)$ from the least chi-square method.
See also extended data in \textbf{Supplementary Note 5, Supplementary Figure 6 and 7}.

If we directly drive the transmon using a separate charge-line, instead indirectly drive through the resonator, the DID can be suppressed approximately by a factor of $\mathscr{g}/\Delta_{qr}$ \cite{DID}. For the system in the dispersive coupling regime, $\mathscr{g}/\Delta_{gr} \ll 1$ is satisfied. Hence, the DID can be significantly reduced.
Since $\Gamma_{\phi,\textbf{DID}}^{q}$ scale linearly with $\Gamma_1^{r}(\omega_d)$, the DID becomes negligible for high-coherence resonators when $\Gamma_1^{r}(\omega)$ is negligible around $\omega \sim \omega_d$. 
For readout resonators that need sufficient external couplings to the feedlines for high readout efficiencies, one can engineer the interface between resonators and feedlines suppressing $\Gamma_1^{r}(\omega_d)$ while keep large enough $\Gamma_1^{r}(\omega_r^g)$, as a similar strategy is used for Purcell filters.

The magnitude of the DID when we tune the Lamb shift to zero is approximately 1 MHz. We can suppress this to 1 kHz with $\mathscr{g}/\Delta_{gr} = 0.1$ and $\Gamma_1^{r}(\omega_d)=10$ kHz, which are achievable values in typical circuit QED experiments.
Nonetheless, it is undeniable that the suggested measures do not thoroughly eliminate the DID and complicate the circuit design.
Therefore, our scheme might not be practical when a superconducting qubit of extremely low pure dephasing rate less than 1 kHz is required.
However, our approach is still available for the other applications where moderate coherence times are acceptable.

\section*{Conclusion}
To summarize, we experimentally realize a large tuning of the Lamb shift $\sim$ 30 MHz with drive strength while minimizing undesired renormalization of the other properties of the transmon-resonator system. We show that the Lamb shift can be engineered even to zero. Our observation is consistent with multi-level transmon spectroscopy as well as other renormalized quantities such as cross-nonlinearities and resonator frequency pulling. The observation also agrees excellently with Floquet theory.

Controlling the Lamb shift could provide more flexibilities in engineering the transition frequencies of superconducting qubits.
The feasibility of tuning the Lamb shift to zero possesses other practical implications.
Our approach can also be implemented to multi-qubit device without substantial complexities.
We provide specific application examples using the above merits in \textbf{Supplementary Note 4}.

\section*{Methods}
\subsubsection*{Eigenenergy calculation}
In this work, we utilize QuTiP to apply Floquet theory to the driven Hamiltonian models presented in the main text. Our goal is to find the quasi-eigenenergies of the driven Hamiltonians ($\widetilde{E}_{n,\alpha_n}$) that are adiabatically connected to the eigenenergies of the undriven Hamiltonians (${E}_{n}$) when the drive amplitudes are turned off ($\Omega_{ge} \rightarrow 0$). We use the `floquet modes' method of QuTiP, which returns the quasi-eigenenergies in the first Floquet Brillouin zone of the given Hamiltonian, i.e., $\widetilde{E}_{n,0}$ for all $n$. However, these values are not sequentially arranged with respect to $n$, and the sequence even changes as $\Omega_{ge}$ varies. Therefore, we need to take additional steps to find the proper Floquet mode number $\alpha_n$ and quasi-eigenenergies. We gradually increase $\Omega_{ge}$ with a sufficiently small step size and, at every step, find the proper $\widetilde{E}_{n,0}$ and corresponding $\alpha_n$ such that they are adiabatically connected to the values obtained in the previous step. At the beginning, when $\Omega_{ge}=0$ is satisfied, we can find ${E}_{n}$ using the `eigenenergies' method without Floquet theory, and therefore finding the proper mode numbers is unnecessary. We properly adjust the step size when increasing $\Omega_{ge}$ to balance accuracy and computation time.

\subsubsection*{Device fabrication and measurement}
The device and cryogenic setup used in this work are identical to those in our previous work \cite{ann_2022}.
The device consists of a transmon coupled to two coplanar waveguide resonators, but only one of the resonators is used in this work because the other one is weakly coupled with a cross-nonlinearity of less than 100 kHz, and therefore not effective in the experiments. 
The transmon and resonators are defined on a 100 nm niobium titanium nitride (NbTiN) film on a 525 $\mu m$ thick silicon substrate \cite{SRON}.
The Al-AlOx-Al Josephson junction of the transmon is fabricated by typical double-angle shadow evaporation.
The device is mounted on the mixing chamber plate of a dilution fridge (LD-400) and shielded from radiation and magnetic field using Cooper and Aluminum cans.
The optical microscope image of the device is presented in \cite{ann_2022}.

\section*{Author contributions}
B.A conceived the study, made the theoretical description, and fabricated the device. B.A also performed the numerical and experimental study. The measurement infrastructure is constructed by G.A.S. B.A. and G.A.S analyzed data. B.A. wrote the manuscript with input from G.A.S.

\section*{Competing interests}
The authors declare no competing interests.

\section*{Data availability}
Data supporting the plots within the main text of this paper are available through Zenodo at http://10.5281/zenodo.7847837.
Further information is available from the corresponding author upon reasonable request.

\section*{Code availability}
Code used to produce the plots within this paper is available from the corresponding author upon reasonable request.

\begin{acknowledgements}
We thank David Thoen and Jochem Baselmans for providing us with NbTiN films. 
Byoung-moo Ann acknowledges support from the European Union’s Horizon 2020 research and innovation program under the Marie Sklodowska-Curie grant agreement No. 722923 (OMT). This work was supported by the National Research Foundation of Korea(NRF) grant funded by the Korea government(MSIT)(RS-2023-00213037).
This work was also supported by Korea Research Institute of Standards and Science (KRISS-GP2024-0013-06).

\end{acknowledgements}

\widetext
\clearpage

\renewcommand{\theequation}{S.\arabic{equation}}
\renewcommand{\thepage}{S\arabic{page}} 
\renewcommand{\thesection}{S\arabic{section}}  
\renewcommand{\thetable}{S\arabic{table}}  
\renewcommand{\thefigure}{S\arabic{figure}}
\setcounter{figure}{0}
\setcounter{equation}{0}

\section*{Supplementary Information}

\subsection*{Supplementary Note 1 : System parameters and descriptions}
Please see Tab.~\ref{tab1} for essential system parameters and descriptions used in the main text. The values of some system parameters invariable during each cooldown are presented in Tab.~\ref{tab2}.

\subsection*{Supplementary Note 2 : Theoretical descriptions}

\subsubsection*{Renormalized interaction – general multi-level systems}
The Hamiltonian of a general driven multi-level system is expressed by
\begin{equation}
\begin{split}
\label{eq:H_1-0}
   &\hat{H}_0(t) = \sum_{n}\omega_n\ket{n}\bra{n}_q+\sum_{n,m}\Omega_{nm}\ket{n}\bra{m}\cos(\omega_d t).
\end{split}
\end{equation}
Here, $\ket{n}$ refers to the eigenstates of the generalized undriven multi-level system.
For both multi- and two-level systems, the dynamics under far off-resonant drives cannot be properly described by the rotating wave approximation (RWA) \cite{sm-rwa-1, sm-rwa-2}.
For time periodical drives, however, we are allow to apply the Floquet theory to calculate the dynamics for arbitrary drive amplitudes $\Omega_{nm}$ without the RWA.
In Floquet formalism, quasi-eigenstates and corresponding quasi-eigenenergies for $\hat{H}_0(t)$ can be expressed by $\widetilde{\ket{n,\alpha}}$ = $e^{i\alpha\omega_d t}\widetilde{\ket{n,0}}$ and $\widetilde{\omega}_{n,\alpha} = \widetilde{\omega}_{n,0} + \alpha\omega_d$, respectively.
Here, $\widetilde{\ket{n,\alpha}}$ is a Floquet mode with an order of $\alpha$.
We also define $\widetilde{\ket{n}} \in \left \{\widetilde{\ket{n,\alpha}} \right\}$ and $\widetilde{\omega}_{n} \in \left \{\widetilde{\omega}_{n,\alpha}\right\}$ that is adiabatically connected to $\ket{n}$ and $\omega_n$ with $\Omega_{nm} \rightarrow 0$. The selected Floquet mode $\widetilde{\ket{n}}$ can be decomposed like
\begin{equation}
\begin{split}
\label{eq:H_1-1}
    \widetilde{\ket{n}} = \sum_{k}\ket{n}^{(k)} e^{-ik\omega_d t},
\end{split}
\end{equation}
and here $\ket{n}^{(k)}$ is a Fourier component of $ \widetilde{\ket{n}}$ at frequency $k\omega_d$, which also can be decomposed into eigenbases of the undriven system $\ket{j}$ like
\begin{equation}
\begin{split}
\label{eq:H_1-2}
    \ket{n}^{(k)} = \sum_{j}c_{n,j}^{(k)}\ket{j}.
\end{split}
\end{equation}
Here, $c_{n,j}^{(k)}$ are time-independent complex numbers.
Let us introduce a time-dependent unitary operator $\hat{U}(t)$ satisfying $\widetilde{\ket{n}}=\hat{U}(t)\ket{n}$. Under this transformation, $\hat{H}_0$ is transformed to
\begin{equation}
\begin{split}
\label{eq:H_2}
   &\hat{\widetilde{H}}_0 = \hat{U}(t)[\hat{H}_0-i\partial/\partial{t}]\hat{U}^{\dagger}(t)
  = \sum_{n}\widetilde{\omega}_{n}\ket{n}\bra{n}.
\end{split}
\end{equation}
In addition, the interaction between the general multi-level system and a single mode resonator is given by
\begin{equation}
\begin{split}
\label{eq:H_1}
    \hat{{H}}_{I} = i\sum_{n,m}{\mathscr{g}}_{nm}\ket{n}\bra{m}(\hat{a} - \hat{a}^\dagger).
\end{split}
\end{equation}
On the transformed basis, the interaction Hamiltonian $\hat{H}_{I}$ can be given by replacing $\ket{n}$ with $\widetilde{\ket{n}}$.
Then, we can express the interaction term by 
\begin{equation}
\begin{split}
\label{eq:H_3}
   &\hat{\widetilde{H}}_{I} = i\sum_{j,j'}\sum_{n,m,k,k'}\mathscr{g}_{nm}c_{m,j'}^{(k')*}c_{n,j}^{(k)}e^{i(k'-k)\omega_d t}\ket{j}\bra{j'}(\hat{a} - \hat{a}^\dagger).
\end{split}
\end{equation}
Exchanging $n\rightarrow j$ and $m\rightarrow j'$, we can define the time-dependent renormalized interaction matrix  $\widetilde{\mathscr{g}}_{nm}(t)$ given by  
\begin{equation}
\begin{split}
\label{eq:H_4}
   &\widetilde{\mathscr{g}}_{nm}(t) = \sum_{k,k',j,j'}\mathscr{g}_{jj'}c_{j',m}^{(k')*}c_{j,n}^{(k)}e^{i(k'-k)\omega_d t}.
\end{split}
\end{equation}
Longitudinal coupling terms appear when $n=m$ is satisfied. Lamb shift control scheme applied to transmons can be also available for the cases where these terms dominate the interaction Hamiltonian.
Thus, we do not need to confine ourselves to a transmon case specifically.
Generally, $|k'-k|$ does not need to be a unity. Specifically for the longitudinal coupling parts ($n=m$) of transmons, however, the terms with $|k'-k|=1$ have negligible magnitudes in our parameter regime. Hence we only take $|k'-k|=1$ terms into consideration in DLC.

\begin{center}
\begin{table}
 \begin{tabular}{||c || c||} 
 \hline
 Symbols & Descriptions \\ [0.5ex] 
 \hline\hline
 $\omega_d$ & drive frequency\\ 
 \hline
 $\omega_r$ & resonator bare frequency\\ 
 \hline
 $\widetilde{\omega}_{nm}$ (${\omega}_{nm}$) & transition frequency of driven (undriven) bare transmon\\ 
 \hline
 $\widetilde{\omega}_{nm}^k$ (${\omega}_{nm}^k$) & transition frequency of driven (undriven) transmon coupled to resonator\\ 
  \hline
 $\widetilde{\omega}_{r}^l$ (${\omega}_{r}^l$) & transition frequency of driven (undriven) resonator coupled to transmon\\ 
 \hline
 $\mathscr{g}$ & transmon–resonator coupling strength \\ 
 \hline
 $\widetilde{\mathscr{g}}_{nm}$ ($\mathscr{g}_{nm}$) & renormalized (bare) transmon–resonator coupling matrix \\ 
 \hline
 $\widetilde{L}_{nm}$ (${L}_{nm}$) & Lamb shift of driven (undriven) transmon \\ 
 \hline
 $\widetilde{P}$ (${P}$) & resonator frequency pulling of driven (undriven) resonator \\ 
 \hline
 $\widetilde{A}$ ($A$) & self-nonlinearity of driven (undriven) transmon \\ 
 \hline
 $\widetilde{\chi}$ (${\chi}$) & cross-nonlinearity of driven (undriven) transmon–resonator system \\ 
 \hline
\end{tabular}
\caption{\label{tab1} Symbols and descriptions.}
\end{table}
\end{center}

\begin{center}
\begin{table}
 \begin{tabular}{||c || c||} 
 \hline
 Symbols & Values (1st, 2nd) \\ [0.5ex] 
 \hline\hline
 $\omega_r/2\pi$ &  $4.335, ~ 4.335$ GHz. \\ 
 \hline
 $\omega_{ge}^{0}/2\pi$ & $5.901, ~ 5.867$ GHz.\\ 
 \hline
 $\omega_{ef}^{0}/2\pi$ & $5.749, ~ 5.715$ GHz.\\ 
  \hline
 $\omega_{fd}^{0}/2\pi$ & $5.587, ~ 5.553$ GHz.\\ 
 \hline
 $\omega_{ge}/2\pi$ & $5.869, ~  5.835$ GHz.\\ 
 \hline
 $\omega_{ef}/2\pi$ & $5.708, ~  5.676$ GHz.\\ 
  \hline
 $\omega_{fd}/2\pi$ & $5.539, ~  5.510$ GHz.\\ 
 \hline
 $\mathscr{g}/2\pi$ &  $248, ~ 245$ MHz.\\ 
 \hline
 $A$  & $152,~ 150$ MHz. \\ 
 \hline
 $\chi$ & $ 5.8, ~ 6.0 $ MHz. \\ 
 \hline
\end{tabular}
\caption{\label{tab2} Values of some system parameters fixed during each cooldown. The values observed or extracted from 1st and 2nd cooldowns are separately presented.}
\end{table}
\end{center}

\subsubsection*{Renormalized transmon–resonator interaction}
A driven transmon is a specific case of Eq.~\ref{eq:H_1-0}, where $\mathscr{g}_{nm}$ and $\Omega_{nm}$ satisfy the below relation.
\begin{eqnarray}
\begin{split}
\label{eq:2-1}
  \mathscr{g}_{nn\pm1} &\approx \mathscr{g}\sqrt{n+1},\\
  \mathscr{g}_{nn\pm l} &\approx 0,\\
  \Omega_{nn\pm1} &\approx \Omega_{d}\sqrt{n\pm1},\\
  \Omega_{nn\pm l} &\approx 0,
\end{split}
\end{eqnarray}
with $l>1$.
For transmons under monochromatic transverse drive fields with sufficient detunings and moderate drive amplitudes as in our case, we have
\begin{equation}
\begin{split}
\label{eq:2-2}
   c_{n,j}^{(k)} \approx 0
\end{split}
\end{equation}
for $k\pm n \neq j$.
Therefore, only the components that meet $k'-k = n-m\pm1$ in Eq.~\ref{eq:H_4} are dominant for transmons, and then the $\widetilde{\mathscr{g}}_{nm}(t)$ can be simplified by 
\begin{equation}
\begin{split}
\label{eq:2-3}
   &\widetilde{\mathscr{g}}_{nm}(t) =\sum_{j,j'} \sum_{{\begin{subarray}{l} k'-k=\\n-m\pm 1\end{subarray}}} \mathscr{g}_{jj'}c_{n,j}^{(k')*}c_{m,j'}^{(k)}e^{i(n-m\pm1)\omega_d t}.
\end{split}
\end{equation}
This simplification can also be justified without using Floquet formalism as presented in our previous works \cite{ann_2022_sm}.
Dropping oscillating terms, we define the time-independent renormalized interaction matrix $\widetilde{\mathscr{g}}_{nm}$ 
\begin{equation}
\begin{split}
\label{eq:2-4}
   &\widetilde{\mathscr{g}}_{nm} =\sum_{j,j'} \sum_{{\begin{subarray}{l} k'-k=\\n-m\pm 1\end{subarray}}} \mathscr{g}_{jj'}c_{n,j}^{(k')*}c_{m,j'}^{(k)}.
\end{split}
\end{equation}
$c_{n,j}^{(k)}$ can be found using Floquet theory in general.
For weakly anharmonic systems, however, we can bypass using Floquet theory while not confining ourselves in the conventional approximation regimes.
We can calculate $\widetilde{\mathscr{g}}_{nm}$ without Floquet theory as presented in our previous work \cite{ann_2022_sm}.
For the calculation presented in Fig.1 of the main text, therefore, we use the approach in \cite{ann_2022_sm}, not using Floquet theory to avoid the hardships of finding the proper Floquet mode numbers.

\subsection*{Supplementary Note 3 : Experimental descriptions}

\subsubsection*{Measurement}


Two types of experiments are performed in this work: two-tone spectroscopy using a 4-port vector network analyzer (Keysight N5222A) and pulsed spectroscopy using a Quantum Machines OPX. A signal generator (Keysight N5183B) is used to apply drive fields to the transmon.
For pulsed spectroscopy, a 20$\mu s$ transmon excitation pulse is followed by a 2$\mu s$ readout pulse. The transmon excitation pulse is up-converted to RF band using IQ modulation mode of a Rohde-Schwarz SGS100A.
The readout pulses are up/down-converted to RF/IF band by IQ mixers. Local oscillator signals of IQ mixers are provided by Rohde-Schwarz SGS100A and Keysight E8257D.

The data in Fig.2, Fig.S1, and Fig.S3 are obtained through pulsed spectroscopy, while the others are acquired through two-tone spectroscopy using the 4-port vector network analyzer. 
Pulsed spectroscopy is preferred for multi-photon transitions as it generally gives signals with better contrasts, but there is a risk of imperfect mixer calibrations. 
To avoid this, mixer calibration is performed from time to time during the experiments.

\subsubsection*{Multi-level transmon spectroscopy}

\begin{figure}
    \centering
    \includegraphics[width=0.6\columnwidth]{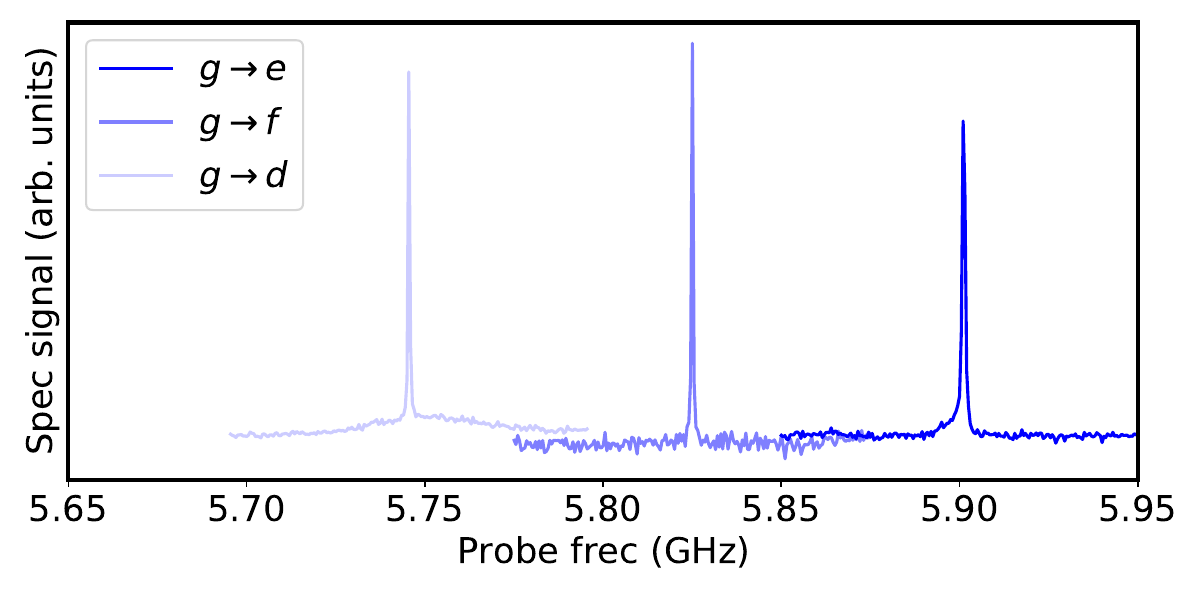}
    \caption{Wideband multi-level spectroscopy of the undriven transmon used in this work. Here, we identify the resonance frequency of $g \rightarrow e$, $g \rightarrow f$, and $g \rightarrow d$ transitions of the transmon.}
    \label{fig_wide}
\end{figure}

Fig.~\ref{fig_wide} shows the experimental result of wideband multi-level spectroscopy to find the transmon's undriven energy levels.
We probe the transmon's second and third excited energy levels by inducing two-photon $g \rightarrow f$ and three-photon $g \rightarrow d$ transitions, respectively. 
Once we find the peaks without driving as in Fig.~\ref{fig_wide}, we trace the resonant frequencies as increasing the power of the driving field.
In this step, we sweep the probe around expected transition frequencies with much narrower scanning ranges down to a few MHz.
The calibrated probe field amplitudes in the frequency $\Omega_p$ for $g \rightarrow e$, $g \rightarrow f$, and $g \rightarrow d$ transitions are $2\pi\times155$ kHz, $2\pi\times6.92$ MHz, and $2\pi\times13.86$ MHz, respectively.
We determine the proper probe field amplitudes for the multi-photon transitions by compromising the efficiency of the measurement and avoiding undesired sideband transitions between the transmon and resonator.

\begin{figure}
    \centering
    \includegraphics[width=0.6\columnwidth]{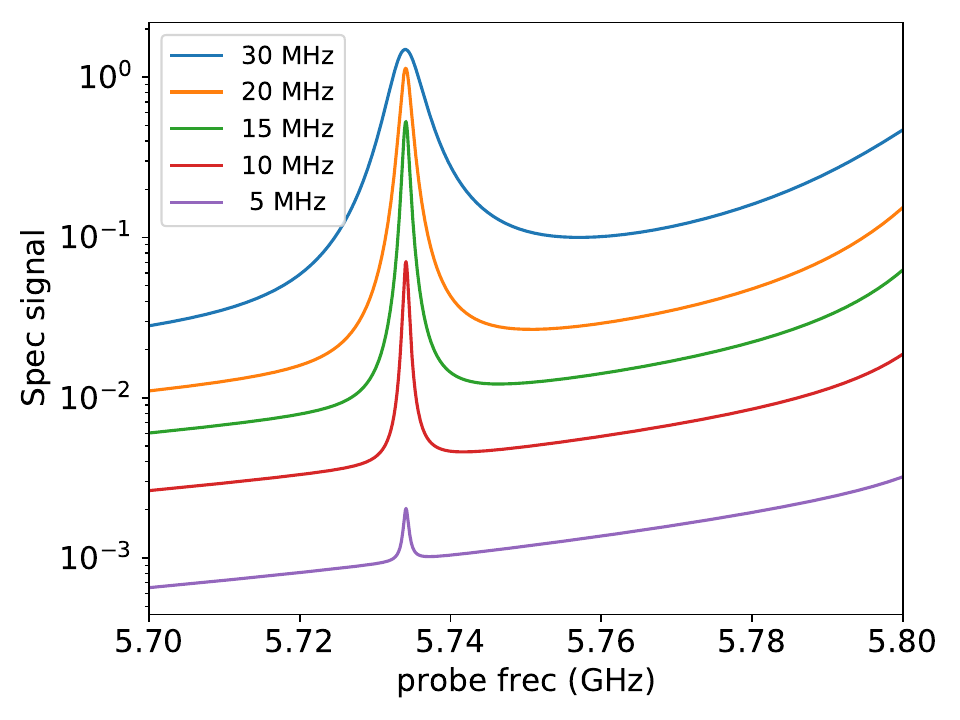}
    \caption{Numerical simulation of probe power effect in $ g\rightarrow d$ transition spectroscopy. We sweep the probe amplitude from 5 to 30 MHz in the simulation while sweeping the probe frequency near the $g\rightarrow d$ transition. The locations of the peaks are nearly invariant. The changes in the backgrounds are the effect of $ g\rightarrow f$ transition. The effect of undesired sideband transitions between the resonator and transmon is not considered in the simulation.}
    \label{fig_probe}
\end{figure}

The probe field alone has negligible impacts on the experiment unless the frequency is very close to the matching conditions for the sideband transitions between the resonator and transmon. The frequency of the probe field is close to the resonant frequency $\omega_q$ of the transmon, and therefore typically far from the matching frequencies for sideband transitions.
See Fig.~\ref{fig_probe} for the numerical simulation that supports our statement. We simulate the probe power effect on $ g\rightarrow d$ transition spectroscopy when only a transmon is presented.

We should also steer clear of scenarios in which certain combinations of the probe and drive frequencies satisfy undesired frequency matching conditions for sideband transitions. 
Such situations can cause discrepancies between theoretical predictions and experimental results. Fortunately, our experimental parameters are far away from such conditions.

\subsubsection*{Discussion on theory–experiment mismatches}
One can identify irregular fluctuations of the experimental data in Fig. 2(c-d) of the main text. We attribute these to the slow drift of the transmon's frequencies during the measurement. We cannot use high probe powers for spectroscopy to minimize the undesired couplings induced by probe and drive tones. The poor contrast in the spectrum due to the small probe tone forces us to take averaging for many hours.

For the disagreement between solid and dashed lines at large $\Omega_{d}$ in Fig 3(d), we strongly suspect that three-photon sideband transition $\ket{d}_q\ket{0}_r$ $\leftrightarrow$ $\ket{g}_q\ket{1}_r$ accounts. $\omega_d=2\pi\times4.2$ GHz is very closed to the matching condition of that transition.
Here, $\ket{\dots}_q$ and $\ket{\dots}_r$ are quantum states of the transmon and resonator, respectively. 
The solid and dashed lines show better agreements for $\omega_d/2\pi=4.0$ and $4.1$ GHz in Fig~\ref{fig_4_ext}.
We can also see disagreement between the experimental data and theories with $\Omega_{ge}/2\pi$ larger than 0.8 GHz. This can be attributed to renormalized couplings among the higher levels of the transmon and resonator, or our model's missing frequency shifts of higher levels due to the unexpected drive-induced couplings to stray modes. 
These are, however, the matter outside of computational subspace, $\left\{ \ket{g}_q,\ket{e}_q \right\}$, and therefore should not be taken into account in the calculation.

\subsection*{Supplementary Note 4 : Possible applications}

In addition to the fundamental interests, our findings enable the efficient tuning of both AC Stark and Lamb shifts simultaneously, which is particularly important for non-parametric frequency control of multi-level systems. The ability to control both AC Stark and Lamb shifts offers greater flexibility. In this section, we will introduce   specific examples of possible applications.

\subsubsection*{Qubit frequency tuning with a global charge line}
\begin{figure}
    \centering
    \includegraphics[width=0.5\columnwidth]{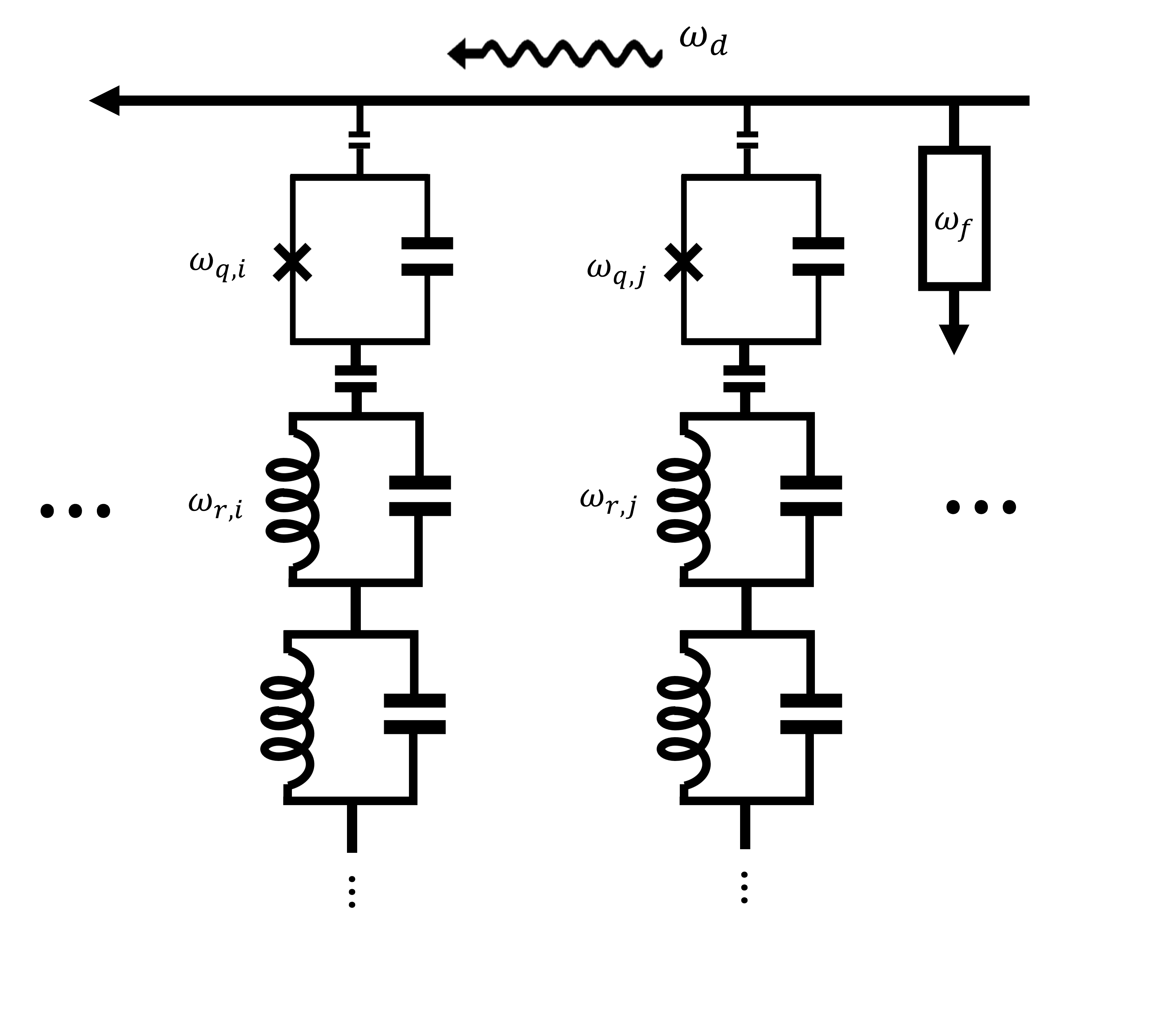}
    \caption{Schematic of qubit frequency tuning with a global charge line . Multiple superconducting qubits ($\omega_{q,n}, n=0,1,...$) coupled to a global charge-line (horizontal straight arrow). The qubits are also dispersively coupled to resonator modes (LC circuits), and we use Lamb shift between qubits and resonator modes as additional degree of freedoms for qubit frequency tuning. Only the frequencies of the fundamental mode are presented ($\omega_{r,n}, n=0,1,...$). We assume that $\omega_{q,n}$ are confined within a finite frequency band. Individual charge-lines to each qubit, couplers among qubits, readout resonators and readout-lines are not presented for simplicity. A rectangular box represents a filter blocking quantum noise near frequency $\omega_f$. We assume $\omega_f \approx \omega_{q,n}$.}
    \label{fig_application} 
\end{figure}

Standard quantum processors based on fixed-frequency superconducting qubits are accompanied by individually assigned charge-lines to each qubit to implement single qubit gates, and readout resonators coupled to readout lines. The most trivial approach of tuning qubit frequencies is inducing AC Stark shift through each charge-lines. However, the couplings to these charge-lines should be small enough not to induce significant energy loss of the qubits. Furthermore, to avoid undesired renormalization in the strong drive limit, sufficient detunings between the qubits and Stark tones are needed. Introducing additional charge-lines to each qubit with sufficiently large couplings and noise filters around qubit frequencies can be a solution, but this approach requires significant overheads. Applying different AC Stark shifts to each qubit using additional global charge-lines, or readout lines is also not easy since normally the frequencies of qubits and resonators are confined within finite band for hardware efficiency.
If some of qubits or readout resonators have similar frequencies, then differently inducing AC Stark shift becomes extremely challenging. The situation becomes more difficult if one wants to have sufficient detunings between the qubits and Stark tones to avoid undesired renormalization.

Our Lamb shift engineering scheme can solve these challenges readily. The proposed circuit design is presented in Fig.~\ref{fig_application}. An array of qubits is coupled to a global charge-line (horizontal arrow). A filter blocking quantum noise near frequency $\omega_f$ is installed. At the same time, qubits are dispersively coupled to the additional resonator modes depicted by LC circuits. We label the frequency of the fundamental modes by $\omega_{r,n}$ ($n=0,1,...$). We also denote the coupling between the qubit and fundamental modes by $\mathscr{g}_{qr,n}$ ($n=0,1,...$). Here, readout resonators, readout-lines, couplers among the qubits, and individual charge-lines are omitted in the schematic. 
Let us assume that the fundamental transmon frequencies $\omega_{q,n}$ ($n=0,1,...$) are similar. By setting $\omega_f\approx\omega_{q,n}$, we can protect qubits from the noise channel made by the global charge-line. Thus, we can have sufficient coupling between the qubits and charge-line.
There is almost no constrain in choosing $\omega_{r,n}$ and $\mathscr{g}_{qr,n}$ as long as they are in the dispersive regime.
If we set $\omega_d$ close to $\omega_{r,i}$, but far-off resonant with $\omega_{r,n}, n\neq i$. Then, only qubit $i$ will undergo different frequency shift due to the renormalization of Lamb shift. One can have more options to choose $\omega_d$ if using the other higher-order modes (LC circuits without mentioning frequencies beside).

Based on this approach, one can individually regulate the frequencies of multiple qubits using a single drive line, reducing the design complexities of multi-qubit devices and thereby improving the integrity of superconducting qubits.
One of the greatest merits of this approach is that all the additional resonators do not need to have couplings to the global charge-line and any of other feedlines. Thus, the global charge-line and additional resonator modes can be easily designed without considering Purcell effect and undesired coupling. This dramatically simplifies the realization of circuits. Furthermore, for ordinary circuit QED systems, implementing more resonators does not make any fabrication overhead.

\subsubsection*{Usage with qubit–resonator tunable couplers}
Currently, tunable couplers are actively being investigated to enable more efficient on-demand Hamiltonian engineering \cite{sm-tunable-coupler}. The dispersive coupling between qubits and resonators, currently achieved through static capacitance, could potentially be replaced by tunable couplers in the future. This would allow for complete turning off of the coupling when it is not desired, which would be useful in minimizing unwanted cross-talk among qubits and resonators.

However, this plan will encounter challenges due to the back-action on the qubit frequencies that occur when the interaction is turned on. Turning on the coupling not only produces the desired effects but also leads to changes in the qubit frequencies, even when the resonators are empty due to Lamb shift. Avoiding such back-action solely through controlling the AC Stark shift requires updating the microwave tones for the AC Stark shift whenever the coupling is turned on and off. Furthermore, if the back-action shifts the qubit frequency downward, then a compensating Stark tone should be placed in the straddling regime \cite{sm-Koch-PRA-2007}, which introduces issues of eigenstate mixing and multi-photon transitions in the strong drive limit.

Obviously, our Lamb shift engineering scheme can overcome all the hurdles mentioned above. If one applies a drive tone with the proper power and frequency, as outlined in our paper, it is possible to keep the qubit frequency unchanged when turning the coupling on and off.

\subsubsection*{Efficient engineering transition frequencies of higher energy levels}

Due to the weakly anharmonic nature, a transmon's each transition frequencies undergo simuilar shifts
under far-off resonant drive fields. 
Consequently, the anharmonicities of transmons is relatively more difficult to engineer than fundamental transition frequencies \cite{sm-higherlevel}. To overcome this, one should use the drive frequency near resonance of transmons. However in this manner, the drive fields can induce undesired effects such as eigenstate mixing and multi-photon transitions in the strong drive limit.

By utilizing our Lamb shift engineering scheme, it becomes possible to more efficiently tune the higher energy levels of the transmon, even with far-off resonant drive fields. As shown in figure 3d, the anharmonicity of the transmon undergoes a dramatic change, almost doubling in value. This significant tuning of the anharmonicity is primarily caused by the DLC. The underlying principle behind this lies in the fact that the magnitudes of the DLC terms, denoted as $|\tilde{\mathscr{g}}_{nn}|$, increase with $n$ for transmons. As a result, each energy level of the transmon experiences a different amount of frequency shift induced by the DLC terms.
Therefore, our scheme can also be useful for efficient controls of the higher energy levels of transmons.

\subsection*{Supplementary Note 5 : Extended data}
\begin{figure}
    \centering
    \includegraphics[width=0.5\columnwidth]{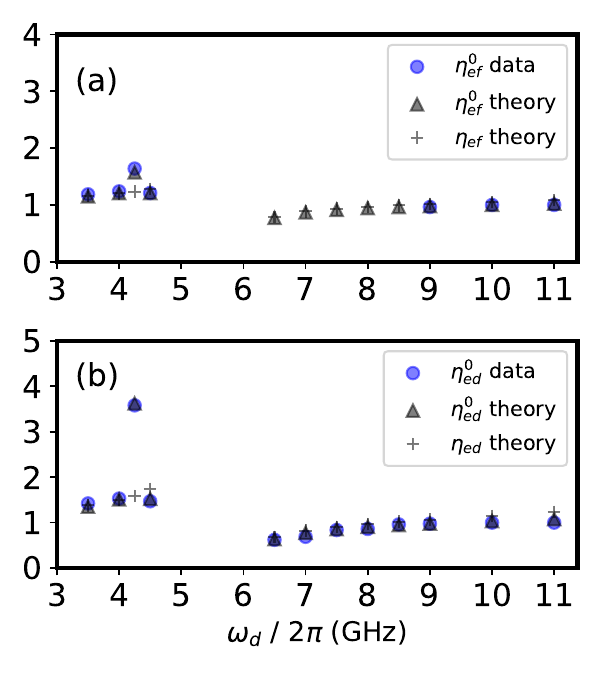}
    \caption{Extended data for figure 2 in the main text. The $g \rightarrow f$ and $g \rightarrow d$ transitions are shown in (a) and (b), respectively. The drive frequency $\omega_d/2\pi$ is swept from 3.5 to 11 GHz. We compare the experimental data (circles) with the theoretical calculation (triangles and crosses, respectively). The theory without the vacuum effect (crosses) fails to capture the data around $\omega_d \approx \omega_r$, clearly indicating the Lamb effect. See the main text for the definition of $\eta$.}
    \label{fig_2_ext}
\end{figure}

\begin{figure}
    \centering
    \includegraphics[width=0.6\columnwidth]{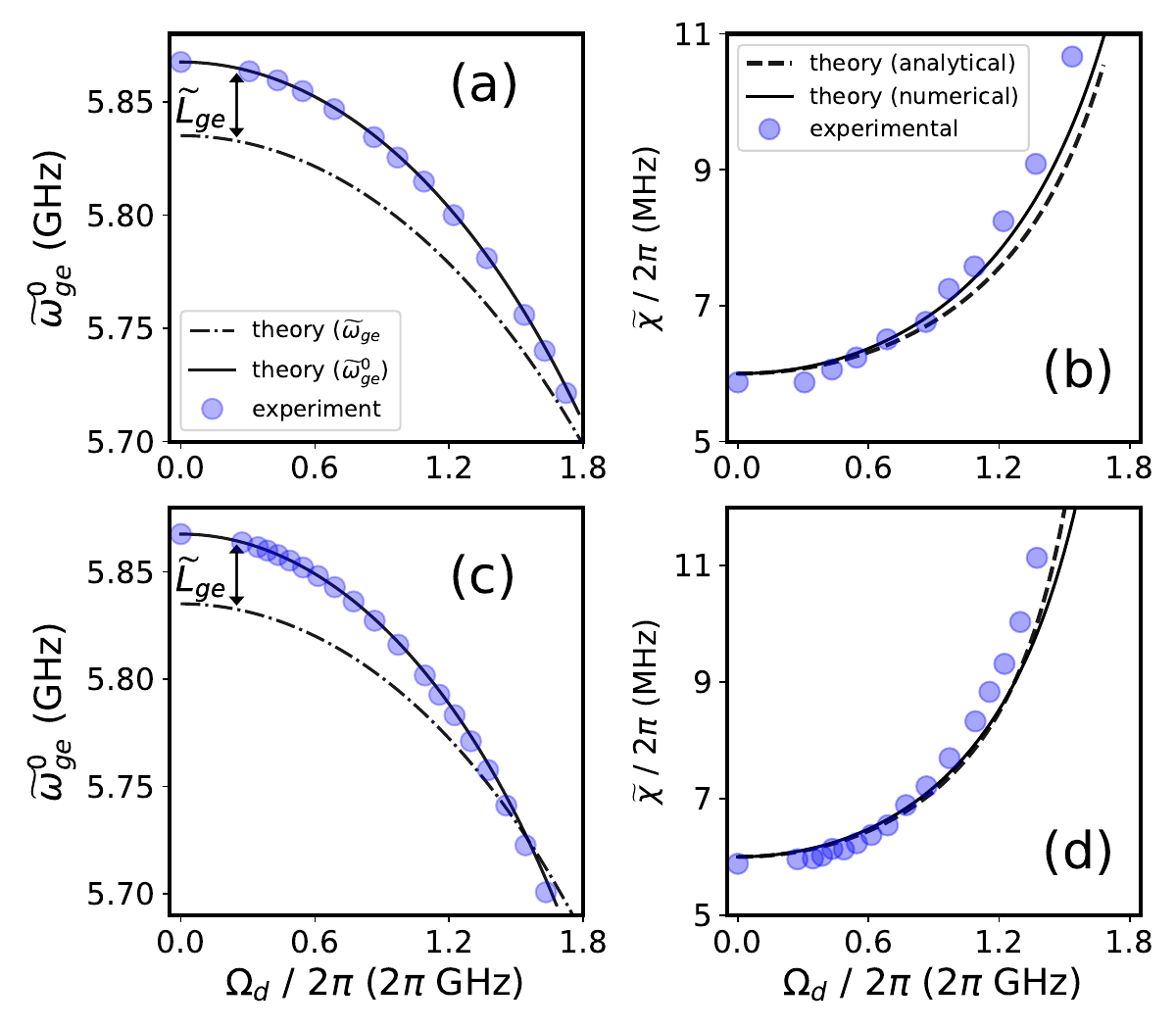}
    \caption{Extended data to figure 3 in main text. We present two cases: $\omega_d/2\pi=4.0$ GHz (a,b) and $\omega_d/2\pi=4.1$ GHz (c,d). In (a,b) and (c,d), one can clearly identify the changes in the Lamb shift and cross-nonlinearity, respectively. The agreements between the solid and dashed lines in (b) and (d) suggest that the renormalization of the transmon-resonator coupling strength is negligible.}
    \label{fig_3_ext}
\end{figure}

\begin{figure}
    \centering
    \includegraphics[width=0.6\columnwidth]{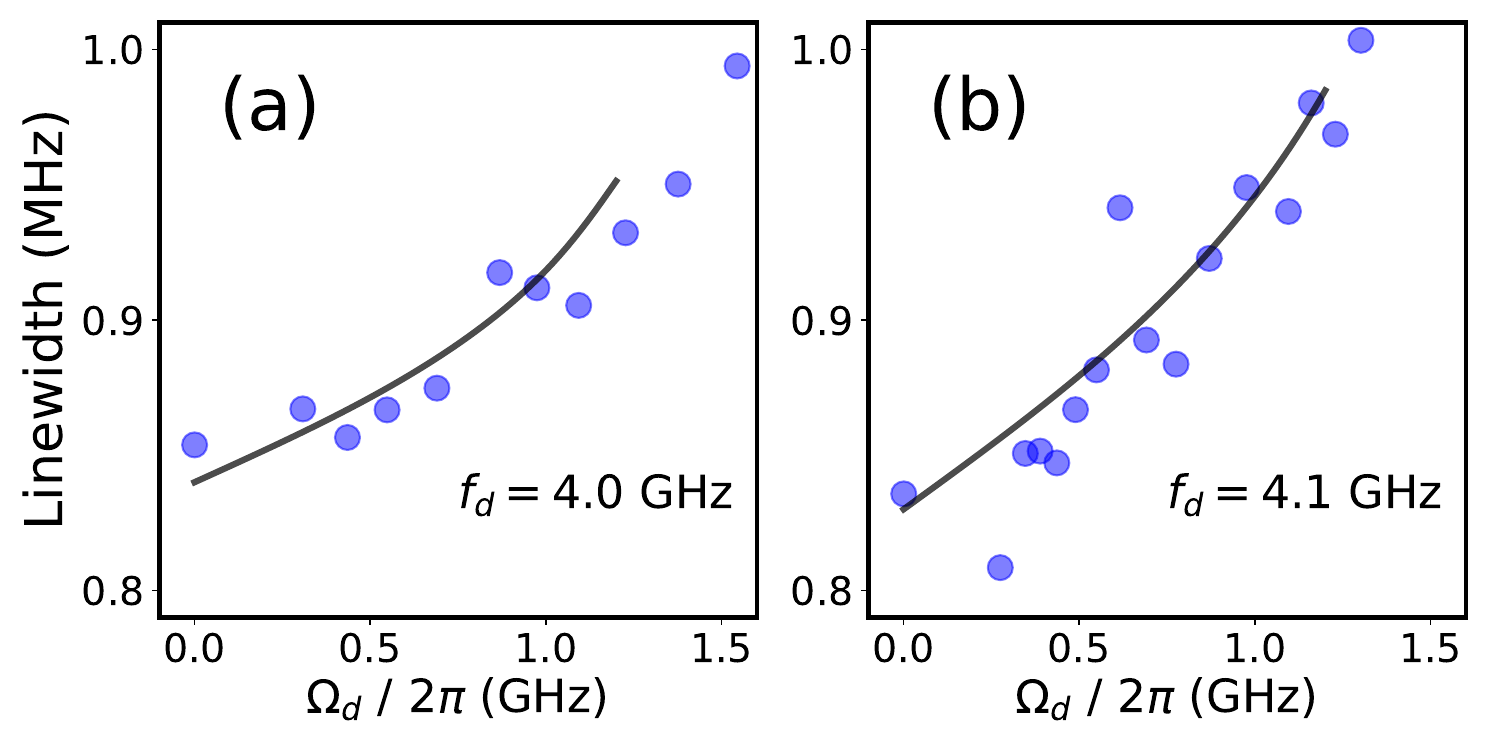}
    \caption{Extended data to figure 4 in main text. We present the cases of $\omega_d/2\pi=f_d=4.0$ GHz (a) and $\omega_d/2\pi=f_d=4.1$ GHz (b). The observed linewidth of the transmon is shown with respect to $\Omega_{ge}$ (circles). The theory plots (lines) are based on Eq.3 in the main text, which show good agreement with the experimental data. The proper values of $\Gamma_1^{r}(\omega_d)$ to quantify other data set with $\omega_d/2\pi=$ 4.0 and 4.1 GHz seems not dramatically different.}
    \label{fig_4_ext}
\end{figure}

\begin{figure}
    \centering
    \includegraphics[width=0.6\columnwidth]{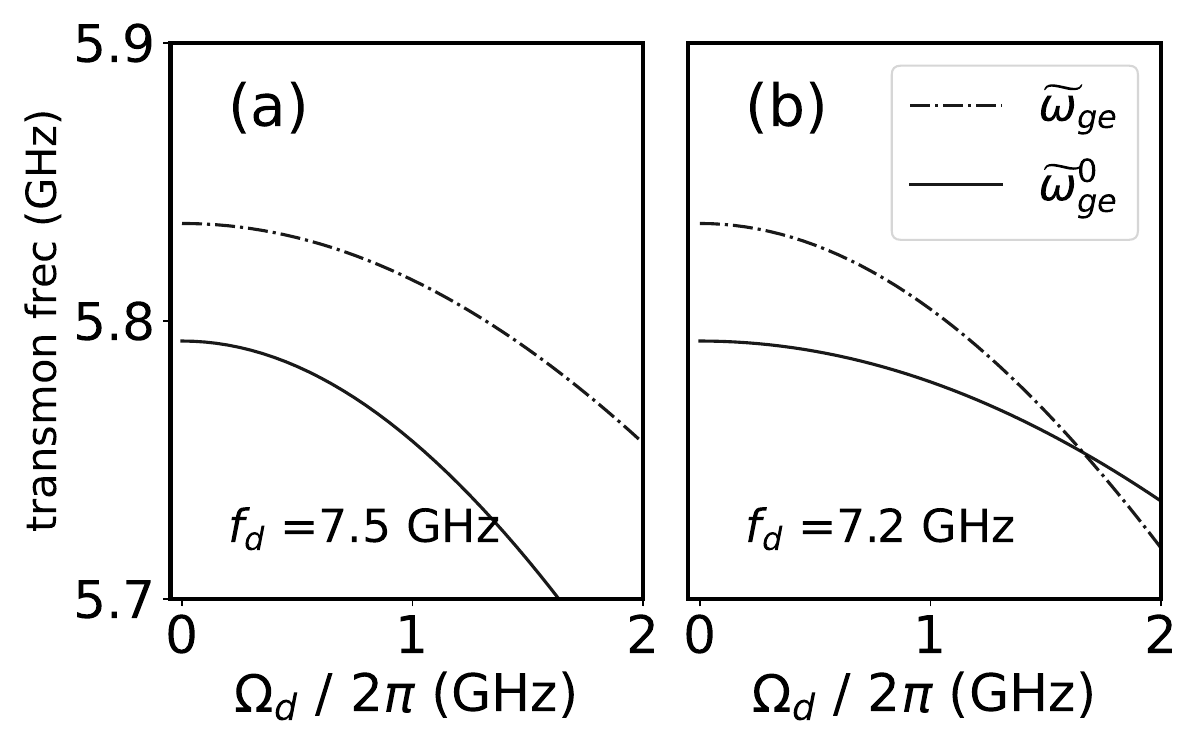}
    \caption{Floquet-based calculation for the case of a transmon coupled with a blue-detuned resonator. 
    In the main text, we only investigated the case of a red-detuned resonator. In this figure, we present numerical studies based on the other configuration. The parameters used for the calculation are identical to the experimental values, except that the resonator frequency is replaced with 7.344 GHz. Depending on the transmon-drive detuning, the Lamb shift varies differently with respect to $\Omega_{d}$. The drive frequency is 7.5 GHz in (a) and 7.2 GHz in (b).}
    \label{fig_blue} 
\end{figure}

Extended data sets to reinforce our arguments given in main text are presented in this section. Please see Fig.~\ref{fig_2_ext}, Fig.~\ref{fig_3_ext}, Fig.~\ref{fig_4_ext}, and Fig.~\ref{fig_blue}.

\subsection*{Supplementary References}

\end{document}